\documentclass[12pt]{iopart}
\usepackage{graphicx}
\usepackage{dcolumn}
\usepackage{bm}

\begin{document}

\title{Fabrication and optical characterization of long-range
  surface-plasmon-polariton waveguides in the NIR}

\author{Markus Weber$^{1,2}$, Johannes Trapp$^1$, Florian B\"ohm$^1$, Peter
  Fischer$^{1}$, Marion Kraus$^1$, Toshiyuki
  Tashima$^{1}$\footnote{Present address: Graduate School of Engineering
    Science, Osaka University, Toyonaka, Osaka 560-8531, Japan}, 
Lars Liebermeister$^1$, Philipp Altpeter$^1$, and Harald Weinfurter$^{1,2}$} 

\address{$^1$ Fakult\"at f\"ur Physik, Ludwig-Maximilians-Universit\"at
  M\"unchen, Schellingstrasse 4, D-80799 M\"unchen, Germany}
\address{$^2$ Max-Planck-Institut f\"ur Quantenoptik, Hans-Kopfermann-Strasse
  1, D-85748 Garching bei M\"unchen, Germany}
\ead{markusweber@lmu.de}

\begin{abstract}
We experimentally demonstrate the propagation of long-range
surface-plasmon-polaritons in a nobel metal stripe waveguide at an optical
wavelength of 780 nm. To minimize propagation damping the lithographically
structured waveguide is produced from a thin gold stripe embedded in a
dielectric polymer. Our waveguide geometry supports a symmetric fundamental
and anti-symmetric first order mode. For the fundamental mode we measure a
propagation loss of $(6.12^{+0.66} _{-0.54})$ dB/mm, in good agreement with
numerical simulations using a vectorial eigenmode solver. Our results are a
promising starting point for coupling fluorescence of individual solid state
quantum emitters to integrated plasmonic waveguide structures.
\end{abstract}

\section{Introduction}
Surface plasmon polaritons (SPP) are electromagnetic (EM) waves guided along
metal-dielectric interfaces. In contrast to free EM waves in homogeneous
dielectric media the hybrid nature of SPP, i.e. collective oscillations of the
metal's free electrons coupled to the EM field in the surrounding dielectric,
provide a way to significantly reduce the effective wavelength. This property
of SPP allows to increase, e.g., the spatial confinement of the propagating
surface wave, opening up the possibility for guiding ``light'' in highly
integrated plasmonic devices \cite{Aussenegg08,Zwiller13}. In addition, the
local enhancement of the electric field at the metal surface
\cite{Lakowicz08,Schell11} opens new possibilities to study and apply
light-matter interaction, e.g. it allows for efficient coupling of
fluorescence light emitted by single quantum emitters to SPP in metallic
nano-structures. In fact, efficient coupling to SPP and preservation of the
quantum nature of the emitted light were demonstrated
\cite{Lukin07,Kolosev09}, however the propagation length of the SPP was
limited to few $\mu$m. A practical way to reduce the strong propagation
attenuation of SPP is to use an only several ten nm thin nobel-metal film (or
stripe) embedded from all sides by the same dielectric
\cite{Berini00,Charbonneau00,Niko03}. Such waveguides support a new class of
coupled SPP modes -- so-called long-range surface-plasmon-polaritons (LRSPP)
\cite{Berini09}-- with a significantly reduced spatial overlap with the lossy
metal and therefore increased propagation length.

First applications of LRSPP waveguides, e.g., the plasmon-assisted
transmission of entangled photons \cite{Woerdman02,Gisin05}, the demonstration
of quadrature-squeezed surface plasmons \cite{Huck09}, and the amplification
of spontaneous emission of many molecules in an LRSPP mode at an optical
wavelength of 600 nm \cite{Gather10} were reported recently. Yet, efficient
optical coupling of single solid state quantum emitters to LRSPP-waveguides
\cite{Leon08} is lacking. As a first step in this direction, in this
contribution we report in detail on the design, fabrication, and optical
characterization of an LRSPP waveguide at an optical wavelength of 780 nm.

\section{Design considerations}

\begin{figure}
\centering
\includegraphics[width=10cm]{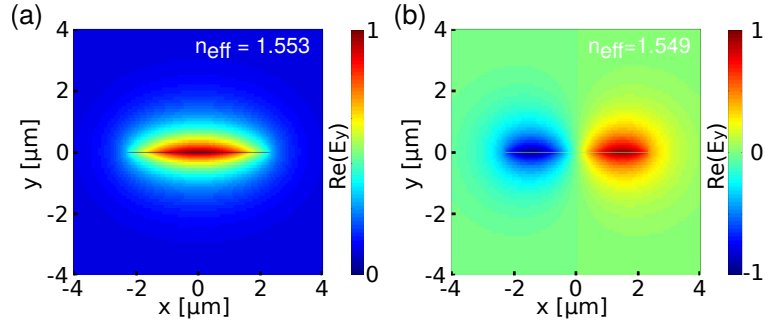}
\caption{Calculated (a) symmetric fundamental and (b) anti-symmetric
  first-order mode of a plasmonic gold-stripe waveguide (width=4.5 $\mu$m;
  thickness=17 nm). Displayed are the real parts of the y-component of the
  electrical field $E$. (Insets: respective effective mode indices)} \label{PlasmonModes:fig1}
\end{figure}

\begin{figure}[tb]
\centering
\includegraphics[width=8cm]{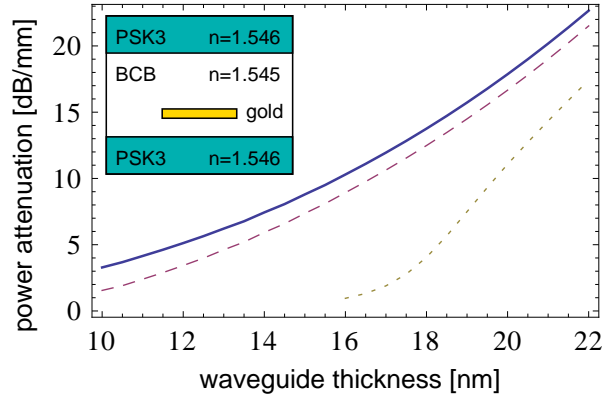}
\caption{Calculated power attenuation of the fundamental (blue solid line),
first-order (cyan dashed line), and the second-order (yellow dotted line)
LRSPP mode as a function of the thickness of a $4.5$ $\mu$m wide gold-stripe
waveguide (at an optical wavelength of 780 nm). The value of the mode power
attenuation is derived from the imaginary part of the respective effective
refractive mode index. (inset: schematic of the waver cross-section, i.e. the
gold waveguide embedded in the polymer BCB which is sandwiched between two
PSK3 microscope slides.)} \label{PowerAttenuation:fig}
\end{figure}

For proper design of the metallic stripe-waveguide, i.e., a quantitative
prediction of the power attenuation as well as of the modal behavior, we first
performed numerical simulations with a vectorial eigenmode solver
\cite{Zhu02}. The width of the gold-stripe (here 4.5 $\mu$m) was chosen to
give good end-fire coupling \cite{Stegeman83, Berini09} efficiency to a
standard single-mode optical fiber. To avoid boundary effects like overlap of
the computed plasmonic modes with the absorbing boundary of the simulation
region we have chosen a $(15 \times 15)$ $\mu$m$^2$ large area of the
surrounding dielectric cladding. This choice of the simulation region
guarantees that the computed power attenuation of each mode, i.e. the
imaginary part of the effective refractive mode index, is only determined by
the spatial overlap of the mode with the cross section of the lossy metal
core. Fig. \ref{PlasmonModes:fig1}(a) and (b) show the simulated symmetric
fundamental and anti-symmetric first-order LRSPP modes for a gold-stripe
waveguide of 4.5 $\mu$m width and 17 nm thickness embedded in the dielectric
polymer benzocyclobutene (BCB) (refractive index: 1.545). Repeating such
simulations for different gold film thicknesses and evaluating the imaginary
part of the effective refractive index of each mode we obtain the desired
power attenuation (see Fig. \ref{PowerAttenuation:fig}). As expected the
attenuation decreases with decreasing gold-layer thickness, independent of the
order of the mode. Additionally, because higher order modes are less localiced
in the gold film, they experience a lower power loss than lower order
modes. 

To summarize our major design criteria, as a general rule of thumb, to
minimize propagation loss of LRSPPs the gold film should be as thin as and the
surrounding dielectric cladding as homogeneous as possible in terms of
refractive index \cite{Berini00,Niko03,Berini09}. From an applied point of
view (to avoid holes and islands during deposition of gold), however, the
minimum propagation loss is limited by a minimum gold layer thicknessis of
around 10 nm.

\section{Fabrication}

Gold-stripe waveguides are fabricated by first spin-coating a 13 $\mu$m thick
layer of BCB on a planar N-PSK3 glass substrate ($15 \times 15$ mm) from
Schott. The BCB layer is baked softly in a drying oven (Memmert UM 100) at
100$^{\circ}$C and 209$^{\circ}$C for a time of 60 s and 40 min,
respectively. The degree of polymerization of BCB after this soft bake is
approximately 80$\%$ \cite{Johannes11}. Onto that, $(13.6 \pm 0.5 )$ nm thick
gold stripes of $4.5$ $\mu$m width are patterned via a lift-off technique. For
that purpose, positive-tone photoresist Microposit S1813 is coated onto the
sample, baked on a temperature stabilized hotplate, exposed with UV light, and
developed in an alkaline developer Microposit 351 (diluted in water 1:3). The
illumination is done with the help of a MJB3 mask aligner (Suess-MicroTec)
using a 200 W mercury arc lamp. Gold is deposited with an electron beam
evaporator (Telemark 271/277-32) which is part of an UHV chamber
(BesTec). Subsequently, removing the photoresist in a bath of
N-Methyl-2-pyrrolidone results in lifting off residual gold areas.

\begin{figure}[tb]
\centering
\includegraphics[width=8cm]{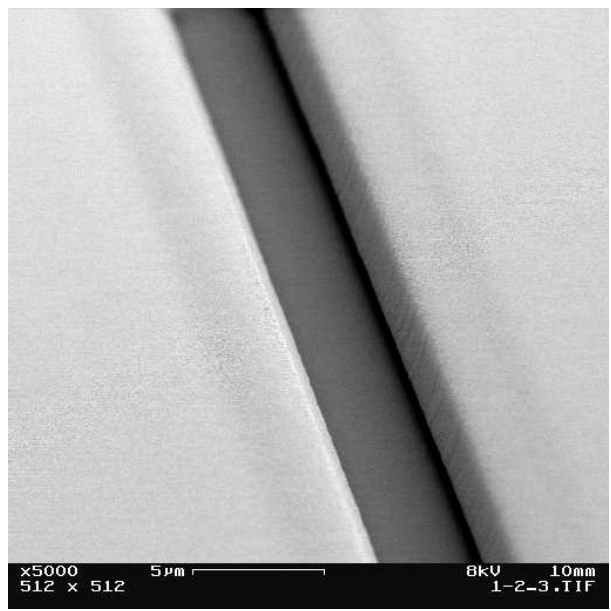}
\caption{Scanning electron microscope image of the sample prior to deposition
of the gold film. Clearly, the undercut in the resist structure is
visible.} \label{REM:figREM}
\end{figure}

In contrast to standard photolithography in our case an adhesion promoter like
Titanium can not be used for deposition of the gold layer due to disturbance
of the refractive index of the sensitive layer structure needed for LRSPP
propagation \cite{Berini00}. For that reason an additional layer below the
photo resist -- a so-called sacrificial layer of lift-off resist (LOR) -- is
used to create an undercut in the resist structure (see
Fig. \ref{REM:figREM}). This additional step is necessary to prevent the
appearance of fence-like lift-off artifacts \cite{Johannes11}. After lift-off
the wafer is coated with a second 35 $\mu$m thick layer of BCB which again is
soft-baked at 100$^{\circ}$C for 60 s. This layer is then sealed with another
planar glass substrate and finally the whole sample is cured in a nitrogen
atmosphere at 233$^{\circ}$C for 120 min.

The quality of the gold stripes are inspected during the fabrication process,
i.e. directly after lift-off. Here, a scanning electron microscope (Zeiss LEO
DSM 982) is used for an overview of defects whereas the stripe width and
height are determined with an atomic force microscope (AFM). To avoid
systematic errors in AFM height measurements on different substrate materials
(here gold and BCB) spare samples from the same fabrication run are coated
prior to inspection with a thin layer of gold \cite{Johannes11,Bar97}. In
addition the thickness of the gold-stripe waveguide is cross-checked
independently from the AFM measurement with an oscillating crystal installed
in the electron beam evaporation chamber. Typical errors in the determination
of the height are on the order of $\pm 0.5$ nm.  For further optical
characterization the samples are first cut with a diamond wire saw to
different lengths (here 1 and 2 mm). Finally, the sample and the waveguide
facets are carefully polished with silicon carbide and aluminum oxide fiber
lapping sheets (minimum grain size $0.3$ $\mu$m).

\section{Optical characterization}
\begin{figure}[tb]
\centering
\includegraphics[width=8cm]{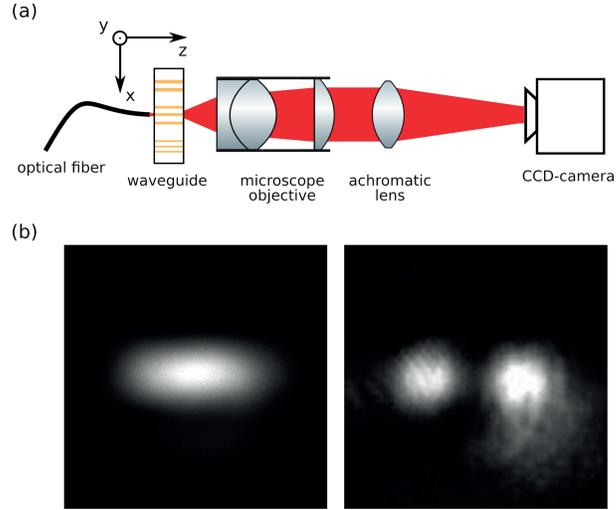}
\caption{(a) Optical setup used to measure the LRSPP mode profiles. (b)
Measured intensity profiles of the symmetric fundamental mode and
superposition with the anti-symmetric first order mode of a 1 mm long stripe
waveguide of $4.5$ $\mu$m width and $(13.5 \pm 0.5)$ height. Depending on the
lateral displacement of the optical fiber used for end-fire coupling either
the fundamental (left) or a superposition with the first order plasmonic mode
is excited (right). Each image displays a $(8\times 8)$ $\mu$m$^2$ region of
the sample facet.} \label{setup:fig}
\end{figure}

To characterize the power attenuation of the LRSPP we analyze the damping of
butt-coupled light (see Fig. \ref{setup:fig}(a)). In detail, a laser beam at a
wavelength of 780 nm is coupled into a polarization maintaining optical single
mode fiber. For optimum end-fire coupling to the LRSPP the cleaved end facet
of the optical fiber is precisely positioned with piezo-controlled mechanical
XYZ-translation stages with respect to the optical axis of the plasmonic
waveguide. To optimize coupling to the $E_y$-polarized LRSPP modes
\cite{Berini00} the exciting laser polarization is aligned parallel to the
y-axis, i.e. perpendicular to the gold layer. For a measured gold stripe width
of $4.5$ $\mu$m and thickness of $(13.5 \pm 0.5)$ nm we numerically estimate
propagation loss of $6.8^{+0.7} _{-0.6}$ dB/mm and a maximum end-fire coupling efficiency of
69 $\%$ for the fundamental LRSPP mode. Light transmitted by the waveguide is
imaged in propagation direction of the LRSPP onto a calibrated CCD camera (see
Fig. \ref{setup:fig}), for detailed analysis of the mode structure.

For a fabricated waveguide of $4.5$ $\mu$m width we observe either the
symmetric fundamental mode or a superposition with the anti-symmetric
first-order LRSPP mode. When the symmetric mode of the butt-coupling fiber is
centered laterally with respect to the waveguide, only the symmetric
fundamental LRSPP mode is excited (see Fig. \ref{PlasmonModes:fig1}(b)). For a
small lateral displacement (less than the mode field diameter) of the optical
fiber with respect to the plasmonic gold-stripe both the fundamental and
first-order LRSPP modes are excited, yet with different excitation
amplitudes. Because these modes exhibit different effective refractive indices
(1.553 and 1.549 at $\lambda=780$ nm for the fundamental and first-order mode,
respectively) they pick up different phases during propagation and at the end
of the waveguide both modes add up coherently, resulting in a multi-mode
interference pattern \cite{Zia06,Lee09}. As such interference pattern depends
sensitively on the ratio of amplitudes and the phase between both LRSPP modes,
experimental analysis of the modal behaviour of our LRSPP waveguides has to
take this into account. E.g., for a large lateral displacement of the
butt-coupling fiber the anti-symmetric first-order LRSPP mode is excited
better than the symmetric fundamental mode. In this case at the output of the
waveguide we observe a multi-mode interference pattern which is dominated by
the anti-symmetric first order mode. As the fundamental LRSPP mode is always
excited this leads to an asymmetry in brightness between the right and left
spot of the observed interference pattern (see Fig. \ref{setup:fig}(b),
right). To clearly demonstrate that the observed modes are of plasmonic nature
the polarization of the exciting laser is rotated by 90 degrees (parallel to
the x-axis). As expected from our numerical simulations of the polarization
properties of the LRSPP modes the observed modes disappear (exctinction ratio
$<$ 1:100) because they can not be excited.

\begin{figure}[tb]
\centering
\includegraphics[width=8cm]{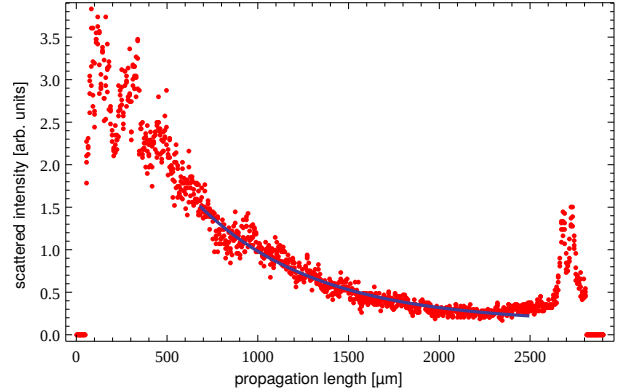}
\caption{Off-scattered light intensity depending on the propagation length
  (measured in top view configuration).} \label{fig:measured_damping}
\end{figure}

To get a representative value of the mode power attenuation we investigate 8
equivalent waveguides on one fabricated chip. Step-by-step we couple laser
light via the butt-coupling fiber to one of the waveguides and optimize the
coupling to the fundamental mode by monitoring the shape of the
outcoupled LRSPP mode at the polished facet of the chip. To determine
the mode power attenuation (propagation damping) of the fundamental LRSPP
mode, we measure with a second calibrated CCD camera (not shown in
Fig. \ref{setup:fig}) the decrease of scattered light in top-view
configuration (optical imaging axis parallel to the y-axis). As the imaged
width of a waveguide spreads over several pixel of the CCD camera, we
integrate the measured intensity of the scattered light over 64 pixel in a
line perpendicular to a waveguide. For each of the 8 waveguides on the chip we
than fit the decrease of scattered light in a line parallel to the waveguide with an
exponential function (see Fig. \ref{fig:measured_damping}). To get an accurate
propagation damping we on purpose skip the in- and out-coupling sections of the
waveguides, where e.g. uncoupled light from the optical fiber leaks into the
sandwitched layer structure of our chip (see e.g. inset of
Fig. \ref{PowerAttenuation:fig}). Averaging over the fit results of all eight
examined waveguides we finally determine a propagation loss of
$6.12^{+0.66}_{-0.54}$ dB/mm. Within our measurement error this finding is in
good agreement with the calculated propagation loss of $6.8^{+0.7}_{-0.6}$ dB/mm.

\section{Summary and outlook}
In summary, we demonstrated the excitation and long-range propagation of a SPP
along a planar gold-stripe waveguide at an optical wavelength of 780 nm. For a
stripe width of 4.5 $\mu$m and height of $(13.5 \pm 0.5)$ nm we measured a
propagation damping of $6.12^{+0.66}_{-0.54}$ dB/mm. This achievement together
with the possibility to embed e.g. narrowband solid state quantum emitters
like SiV centers hosted in diamond nano-crystals \cite{Neu11} in the BCB
cladding layer \cite{Gather10,Schell12} forms an ideal basis for future
efficient coupling of single-emitter resonance fluorescence to LRSPP
waveguides.

\section*{Acknowledgments}
We acknowledge funding from the DFG through the excellence cluster NIM and the
Forschergruppe 1493. TT acknowledges support from the Japanese Society for the
Promotion of Science. 


\end{document}